\documentclass[onecolumn,trackchanges]{aastex63}
\usepackage[caption=false]{subfig}
\usepackage{graphicx}
\usepackage{epstopdf}
\usepackage{color}
\usepackage{multirow}
\usepackage{graphicx}
\usepackage{booktabs}
\usepackage{amsmath,amssymb}
\usepackage[T1]{fontenc}

\graphicspath{{./}{figures/}}

\begin{document}

\title{Do Prompt Gamma-ray Burst Fireball Composition Impact on Afterglow Emission? Cases Study for Long GRBs 080916C/090902B and Short GRBs 090510/130603B}
\correspondingauthor{Xiao-Li Huang}  
\email{xiaoli.huang@gznu.edu.cn}

\author{Yu Gan}
\affiliation{School of Physics and Electronic Science, Guizhou Normal University, Guiyang 550025, People’s Republic of China}

\author[0009-0003-0580-6724]{Ren-Jie Xiong}
\affiliation{Guangxi Key Laboratory for Relativistic Astrophysics, School of Physical Science and Technology, Guangxi University, Nanning 530004, People’s Republic of China}

\author[0009-0001-8025-3205]{Zi-Qi Wang} 
\affiliation{Guangxi Key Laboratory for Relativistic Astrophysics, School of Physical Science and Technology, Guangxi University, Nanning 530004, People’s Republic of China}

\author[0009-0001-0529-1172]{Qi-Yu Yan} 
\affiliation{Guangxi Key Laboratory for Relativistic Astrophysics, School of Physical Science and Technology, Guangxi University, Nanning 530004, People’s Republic of China}

\author{Liang-Jun Chen} 
\affiliation{Guangxi Key Laboratory for Relativistic Astrophysics, School of Physical Science and Technology, Guangxi University, Nanning 530004, People’s Republic of China}

\author[0000-0002-9725-7114]{Xiao-Li Huang}
\affiliation{School of Physics and Electronic Science, Guizhou Normal University, Guiyang 550025, People’s Republic of China}
\affiliation{National Astronomical Observatories, Chinese Academy of Sciences, Beijing 100101, People's Republic of China}

\begin{abstract}
Broadband observations with the {\em Fermi} mission reveal that a large fraction of gamma-ray burst (GRB) spectra are dominated by non-thermal emission, while a small fraction are dominated by thermal/quasi-thermal emission, likely indicating the difference in jet composition among GRB. By selecting two typical long GRBs (080916C and 090902B) and two short GRBs (130603B and 090510), we present a comparative analysis to investigate whether the composition of prompt GRB jets influences the afterglow emission for bursts originating from both massive star collapse and compact binary mergers. 
Incorporating emission from both primary and cascade electron populations, we fit the multi-wavelength afterglow lightcurves of these GRBs with the standard forward shock model and analyze the particle acceleration and radiation physics of the jets. 
Our results show that the afterglow lightcurve evolution with the characteristic parameters is not related to the composition of the GRB fireball but rather depends on the ambient medium density. The early UV-optical afterglows of the two long GRBs are dominated by synchrotron emission from cascaded $e^{\pm}$ pairs produced via the $\gamma\gamma$ annihilation process, whereas this is not the case for the two short GRBs. These results suggest that the internal energy of the fireball is converted into jet kinetic energy during the prompt phase, and that the fireball composition leaves no detectable footprint on the afterglow jet. Instead, the density of the ambient medium plays an essential role in shaping the afterglow emission.   
\end{abstract}


\section{Introduction}
\label{sec:intro}
Gamma-ray bursts (GRBs) are short, intense $\gamma$-ray flashes. They are produced by ultra-relativistic jets launched during the gravitational collapse of massive stars (long GRBs with duration longer than 2 s) or mergers of compact stellar objects (short GRBs with duration shorter than 2 s) \citep{1986ApJ...308L..43P,1989Natur.340..126E,1992ApJ...395L..83N,1999ApJ...524..262M,2004RvMP...76.1143P,2006ARA&A..44..507W,2015PhR...561....1K}. The composition of GRB jets remains an open basic question. Theoretically, GRB jets have been proposed as a ``hot" fireball, in which the outflow is composed of electron-positron plasma, photons, and a small amount of baryons \citep{1986ApJ...308L..47G,1986ApJ...308L..43P,1990ApJ...365L..55S,2002ARA&A..40..137M,2004RvMP...76.1143P}. Particles undergo First-order Fermi acceleration at internal shocks within the ultra-relativistic flow and subsequently cool via synchrotron and inverse Compton processes, producing the observed prompt $\gamma$-ray emission \citep{1993ApJ...405..278M,1994ApJ...430L..93R}. A characteristic prediction of this scenario is the coexistence of a thermal or quasi-thermal photospheric component with a non-thermal component from shock accelerated electrons \citep{1986ApJ...308L..43P,1986ApJ...308L..47G,2000ApJ...530..292M}. Indeed, such thermal signatures have been identified in a subset of GRBs, and in some cases constitute the dominant contribution to the prompt emission spectra \citep{2010ApJ...709L.172R,2010ApJ...716.1178A,2011MNRAS.415.1663T,2012MNRAS.420..468P}. Notably, the spectra of GRBs 090510 and 090902B are dominated by a prominent quasi-thermal component \citep{2009ApJ...706L..33G,2011MNRAS.415.1663T,2011ApJ...730..141Z}, and GRB 220304A similarly exhibits a thermal-dominated prompt emission \citep{2024ApJ...964...45C}. Furthermore, the time-resolved properties of GRB 220426A follow the correlations between the bulk Lorentz factor and the prompt luminosity ($\Gamma-L$), as well as between the spectral peak energy and the luminosity ($E_{p}-L$), consistent with the patterns observed in GRB 090902B \citep{2022ApJ...940..142W}.
Analogously, GRB 231129C exhibits a hard low-energy and soft high-energy spectrum, in line with the quasi-thermal spectral characteristics of GRB 090902B, indicating a substantial thermal contribution \citep{2024ApJ...972..132C}.   

Alternatively, numerous GRB observations favor a Poynting-flux-dominated jet over a thermal-component-dominated model. In this framework, the observed GRB emission is powered by the magnetic energy dissipation within the ejecta \citep{1994MNRAS.270..480T,1994MNRAS.267.1035U,1997ApJ...482L..29M,2003astro.ph.12347L,2003ApJ...596.1080V}. The prompt $\gamma$-rays are then explained as synchrotron radiation from non-thermal electrons accelerated in magnetic reconnection regions within the Poynting-flux-dominated jet \citep{2011ApJ...726...90Z}.
\cite{2009ApJ...703.1696Z} reported that GRB 080916C showed no evidence for the thermal component predicted by baryonic jet models, strongly suggesting that a significant fraction of the outflow energy is likely carried by the Poynting flux. Similarly, \cite{2012ApJ...758L..34Z} analyzed the time-dependent spectrum of GRB 110721A and found it consistent with a non-thermal emission from the optically thin region of the GRB outflow. Broadband observations of GRB 221009A further reveal a distinctive two-component jet structure, consisting of a narrow ($\sim 0.6$ half-opening angle) Poynting-flux-dominated jet and a broader, matter-dominated jet with angular structure \citep{2024JHEAp..41...42Z}. 
Collectively, these observations indicate two distinct types of GRB outflows: baryonic jets and Poynting-flux-dominated jets, featuring prominent thermal or quasi-thermal components and dominant non-thermal synchrotron emission in prompt spectra, respectively. 

Beyond the prompt emission, GRBs are typically followed by long-lasting, multi-wavelength afterglows. Observations across these bands provide insights into jet properties not accessible from the prompt emission alone. How the jet composition influences energy partition, Lorentz factor evolution, and magnetic field strength, and how these shape synchrotron and synchrotron self-Compton (SSC) emission and photon-photon annihilation, remains an open question. Motivated by this, we investigate the afterglow properties and radiation signatures of representative GRBs under both baryonic and Poynting-flux-dominated scenarios.
Specifically, we analyze the radiation physics of four representative GRBs, 080916C, 090902B, 090510, and 130603B, within the framework of the forward shock model. Using the forward shock model, we examine the synchrotron and SSC emission in both homogeneous and wind-like environments, and incorporate cascade emission processes induced by the absorption of high-energy photons. This paper is organized as follows: Section 2 describes our model; Section 3 presents case studies of GRBs 080916C, 090902B, 090510, and 130603B; The results are discussed in Section 4, and a summary is provided in Section 5.

\section{Models}
\label{sec:Lightcurves}
As the prompt GRB jet propagates outward, it eventually interacts with the external circumburst medium at a radius of $\gtrsim 10^{16}\,\rm cm$. This drives a relativistic external shock that accelerates electrons, producing multi-wavelength afterglow emission via synchrotron and SSC processes. In many cases, the early afterglow overlaps with the tail of the prompt emission, rendering the transition between the two phases gradual rather than sharply defined \citep{2018pgrb.book.....Z}.  

The standard external shock afterglow model well explains the multi-wavelength afterglow (e.g.  \citealp{1998ApJ...497L..17S,1999MNRAS.309..513H}). 
The model parameters include the isotropic kinetic energy $E_{\rm k,iso}$, the bulk Lorentz factor $\Gamma$, the density parameter ($n$ for the ISM or $A$ for the wind medium), the jet opening angle $\theta_{j}$, the fraction of shock energy transferred to electrons $\epsilon_e$, the fraction of shock energy transferred to the magnetic field $\epsilon_B$, the electron spectral index $p$. We consider a uniform, ultra-relativistic GRB jet. Within the jet opening angle $\theta_j$, the energy per unit solid angle, bulk Lorentz factor, and matter distribution are uniform, dropping sharply to zero outside this angle. The density profile is expressed as

\begin{equation}
n(t) = 
\begin{cases} 
n_{0}; & \text{for ISM} \\
({\pi m_p c})\, E^{-1}\, A^{2}\, t^{-1};\ & \text{for wind medium},
\end{cases}  
\end{equation}
where $A = \frac{\dot{M}}{4 \pi m_p v} = 3.0 \times 10^{35} A_{*}\,\rm cm^{-1}$ and $A_{*} = \frac{\dot{M}/10^{-5}M_{\odot}\,\rm yr^{-1}}{v/10^{3}\rm \,km\ s^{-1}}$, $ m_p$ is the proton mass, $c$ is the speed of light, $E$ represents the general form of $E_{\rm k,iso}$, $\dot{M}$ is the mass-loss rate of the massive star and $v$ is the constant wind speed for a Wolf–Rayet star \citep{1998MNRAS.298...87D,2000ApJ...536..195C,2000ApJ...543...66P}. 

The dynamics of the external fireball is taken as the same reported by \cite{1998ApJ...497L..17S}, \cite{2000ApJ...543...66P}, and {\cite{2001ApJ...548..787S}.  
The temporal evolution of the radius $R$ is given by

\begin{equation}
R(t) = 
\begin{cases} 
\left( \frac{6}{\pi m_p c} \right)^{1/4} E^{1/4}\, n_{0}^{-1/4}\, t^{1/4}; & \text{for ISM} \\
\left( \frac{1}{\pi m_p c} \right)^{1/2} E^{1/2}\, A^{-1/2}\, t^{1/2}; & \text{for the wind medium}.
\end{cases}  
\end{equation}
The temporal evolution of the jet bulk Lorentz factor is given by 

\begin{equation}
\Gamma(t) = 
\begin{cases} 
\left( \frac{1}{8 c} \right)^{3/8}\left(\frac{3}{4\pi m_p c^{2}} \right)^{1/8} E^{1/8}\, n_{0}^{-1/8}\, t^{-3/8}; & \text{for ISM} \\
\left( \frac{1}{4 \pi m_p c^{2}} \right)^{1/4} E^{1/4}\, A^{-1/4}\, t^{-1/4}; & \text{for the wind medium}.
\label{eq:Gamma}
\end{cases}  
\end{equation}

We consider both the primary and cascade electron populations, which produce radiation via synchrotron and SSC processes. 
The energy distribution of the primary electrons is assumed to follow a single power-law $dN/d\gamma_{e}\propto\gamma_{e}^{-\rm p}$, where $\gamma_{e}$ is the electron Lorentz factor. The injected electron spectrum is characterized by the minimum Lorentz factor $\gamma_m = \frac{p-2}{p-1} \frac{\epsilon_e}{\xi_e} \frac{m_p}{m_e} (\Gamma - 1) + 1$, the cooling Lorentz factor $\gamma_c = \frac{6\pi m_e c}{\sigma_T B'^2 t'}$, and the maximum Lorentz factor $\gamma_{\max} = \left( \frac{6\pi e}{\sigma_T B'} \right)^{1/2}$, where $\xi_e$ is the fraction of electrons accelerated, $m_e$ is the electron mass, $\sigma_T$ is the Thomson scattering cross section.

It is generally believed that the radiation region is optically thick during the very early afterglow stage, leading to photon accumulation within the jet. High-energy photons with energy $\epsilon_{\gamma}$ undergo pair production absorption via interactions with target photons of energy $\epsilon_{t}>\Gamma^{2}(m_{e}c^{2})^{2}/\epsilon_{\gamma}$ within the source. This process initiates an electromagnetic cascade, redistributing the energy of high-energy photons into lower energy photons until the opacity for secondary photons drops below unity ($\tau_{\gamma\gamma} < 1$). \cite{2021ApJ...908..225H} proposed that electromagnetic cascade processes can significantly contribute to the early radio, optical, and X-ray afterglow emissions, a result supported by observations of GRBs 050801, 080310, and 110213A
\citep{2022ApJ...939...39W,2024ApJ...966L..25X}. 
Secondary $e^{\pm}$ pairs produce cascade emission via synchrotron and SSC processes \citep{2005MNRAS.363..954M,2013ApJ...768...54B,2017ApJ...847...39V}.
The Lorentz factors of the electron/positron pair are $\gamma_1=f_{\gamma_{e}} \epsilon$ and $\gamma_2=(1-f_{\gamma_{e}}) \epsilon$, where $f_{\gamma_{e}}$ is the energy fraction of $\varepsilon_{\gamma}$. The photons emitted by the primary electrons constitute the first-generation photon field, denoted by $\dot{N}_{\epsilon}^{0}$. Secondary high-energy photons are produced via synchrotron and SSC processes, with a production rate denoted by $\dot{N}_{\epsilon}^{\rm sec}$. Taking absorption into account, the spectrum of escaping (observable) photons can be calculated as

\begin{equation}
\begin{split}
\dot{N}_{\epsilon}^{\rm esc}=(\dot{N}_{\epsilon}^{0}+\dot{N}_{\epsilon}^{\rm sec})(\frac{1-e^{-\tau_{\gamma\gamma}(\epsilon)}}{\tau_{\gamma\gamma}(\epsilon)}),
\end{split}
\end{equation}
where $\tau_{\gamma\gamma}(\epsilon)$ is the optical depth of photons due to $\gamma\gamma$ absorption.
The production rate of $e^{\pm}$ pairs due to $\gamma\gamma$ absorption can be written as 
$\dot{N}_{\rm e}^{\gamma\gamma}(\gamma_{e})=f_{\rm abs}(\epsilon_{1})(\dot{N}_{\epsilon_{1}}^{0}+\dot{N}_{\epsilon_{1}}^{\rm sec})+f_{\rm abs}(\epsilon_{2})(\dot{N}_{\epsilon_{2}}^{0}+\dot{N}_{\epsilon_{2}}^{\rm sec})$,
where $\epsilon_{1}=\gamma_{e}/f_{\gamma_{e}}$, $\epsilon_{2}=\gamma_{e}/(1-f_{\gamma_{e}})$, and $f_{\rm abs}(\epsilon)\equiv1-(1-e^{-\tau_{\gamma\gamma}
(\epsilon)})/\tau_{\gamma\gamma}(\epsilon)$ is the absorption factor (\citealp{2013ApJ...768...54B,2017ApJ...847...39V}).

Finally, the distribution of the cascaded electron population in a quasi-steady state within a time interval $\delta t$ can be given by

\begin{equation}
\label{eq:Ne1}
N_e^{\rm sec}(\gamma_e,t + \delta t)=N_e^{\rm sec}(\gamma_e^*,t)\frac{d\gamma_e^*}{d\gamma_e} + \left\{
\begin{array}{lll}
\frac{1}{\dot{\gamma}_e}\int_{\gamma_e}^{\infty} d\widetilde{\gamma}_e\dot{N}_e^{\gamma\gamma}(\widetilde{\gamma}_e,t + \delta t), && t_{e}^{\rm cool}(\gamma_{e}) < \delta t,\\
\dot{N}_e^{\gamma\gamma}(\gamma_e,t + \delta t)\delta t, && t_{e}^{\rm cool}(\gamma_{e}) > \delta t.
\end{array}\right.
\end{equation}
Here, $\dot{\gamma_{e}}$ is the energy loss of electrons through synchrotron and SSC processes from both the primary electron population accelerated within the jet and the cascade electron population via $e^{\pm}$ pair production.
$\gamma_e^*$ is the electron Lorentz factor at $t$, and $t_{e}^{\rm cool}$ is the electron cooling timescale. Due to cooling, the Lorentz factor decreases from $\gamma_e^*$ to $\gamma_e$ during the interval $\delta t$. This effect is important as it redistributes energy from high-energy SSC photons to lower-energy photons \citep{2021ApJ...908..225H}. The Klein-Nishina (KN) effect is also considered \citep{2005MNRAS.363..954M,2009ApJ...703..675N,2010ApJ...712.1232W}.

\section{Case Study}
With the incorporation of cascade emission,  we investigate the afterglow emission
properties of both matter-dominated and Poynting-flux-dominated jets in the framework of the external shock models. Two long-duration GRBs (080916C and 090902B) and two short-duration GRBs (090510 and 130603B) are selected as representative samples. We fit the afterglow lightcurves of the long and short GRBs in the wind medium and ISM scenario, respectively. 
Within the framework of the uniform jet model, the jet opening angle can be inferred from the afterglow break time $t_{j}$ and is defined as $\theta_{j}\simeq 1/\Gamma(t_{j})$, with $\Gamma(t_{j})$ given by Equation (\ref{eq:Gamma}) for ISM and the wind medium, respectively \citep{2000ApJ...541L...9K, 2001ApJ...562L..55F}.
The model parameters are constrained with the Markov Chain Monte Carlo (MCMC) method. We employ the affine-invariant ensemble sampler \citep{2010CAMCS...5...65G} implemented in the {\tt emcee} package \citep{2013PASP..125..306F}, adopting flat priors on the logarithm of each free parameter and a likelihood function $\mathcal{L}\propto\exp(-\chi^{2}/2)$. The best-fit values and uncertainties reported in Table~\ref{parameter sets} correspond to the posterior medians and the 16th-84th percentiles of the marginalized posterior distributions, while parameters quoted without uncertainties (e.g., $p$) are held fixed during the fitting.

\subsection{Poynting-flux-dominated Long GRB 080916C and Matter-dominated Long GRB 090902B}

 \begin{itemize}
  \item GRB 080916C. It is a bright long burst with a duration of $\sim 66 \rm \,s$ at a redshift of $z = 4.35$. The emission above 100 MeV was detected by {\em Fermi}/LAT\citep{2009Sci...323.1688A, 2009A&A...498...89G}. {\em Swift}/XRT began observing the source about 17 hours after the Fermi trigger, and optical observations were also obtained post-burst \citep{2009A&A...498...89G, 2011RAA....11.1046F}. The isotropic-equivalent gamma-ray energy during the prompt phase is $E_{\gamma,\rm iso} = 5.7^{+0.54}_{-0.41} \times 10^{54} \rm\, erg$ \citep{2011ApJ...730..141Z}.
  It is considered to host a Poynting-flux-dominated jet. 
  Figure~\ref{LC} presents the observed multi-wavelength afterglow lightcurves, spanning from optical to $\gamma$-ray bands, with the inset showing the prompt $\gamma$-ray lightcurve observed by {\em Fermi}/GBM. 
  We fit the global afterglow lightcurve using our model, which includes both the primary+cascade synchrotron and SSC components. 
  The fitting results are shown in Figure~\ref{LC}, and the model parameters are reported in Table~\ref{parameter sets}. The observed multi-wavelength afterglow emission is well reproduced by our model. 
  
  \item GRB 090902B. This long burst has a duration of 21.9 s and a redshift of $z = 1.822$ \citep{2009ApJ...706L.138A,2009GCN..9873....1C}. Its isotropic-equivalent gamma-ray energy in the prompt phase is $E_{\gamma,\rm iso} = 1.77\pm0.01 \times 10^{52} \rm\, erg$ \citep{2011ApJ...730..141Z}. This burst is believed to host a matter-dominated jet. We analyze the multi-wavelength afterglow lightcurves by incorporating primary and cascade synchrotron processes along with the corresponding SSC emission. The dataset includes {\em Fermi}/LAT (0.1 GeV) data from \cite{2011RAA....11.1046F} and \cite{ 2011ApJ...730..141Z}, the optical data from \cite{2010ApJ...714..799P} and \cite{2011RAA....11.1046F}, the 8.5 GHz radio data from \cite{2011RAA....11.1046F}.
  The fitting results are shown in Figure~\ref{LC}, and the corresponding model parameters are listed in Table~\ref{parameter sets}. The afterglow emission from $\gamma$-ray to radio bands is in good agreement with our theoretical expectations. 
\end{itemize}

\subsection{Poynting-flux-dominated Short GRB 130603B and Matter-dominated Short GRB 090510}

\begin{itemize}
   \item GRB 130603B. This burst is classified as a Poynting-flux-dominated short burst, with a duration of 0.18 s and a redshift of $z = 0.3568 \pm 0.0005$ \citep{2013ApJ...777...94C}. Its isotropic-equivalent gamma-ray energy in the prompt phase is $E_{\gamma,\rm iso} \simeq 2.1 \times 10^{51} \rm\, erg$ \citep{2013GCN.14772....1F}. Radio and optical (r-band) data are taken from \cite{2013ApJ...777...94C} and \cite{2013Natur.500..547T}. Employing the standard dynamical evolution model for GRB afterglows, we analyze the multi-wavelength light curves, incorporating an energy injection of $E_{\rm in} = 1.8 \times 10^{52}~\mathrm{erg}$ at $t_{\tau} = 751 \,\rm s$ from the central engine. 
   Our results show that the observed multi-wavelength afterglow is well reproduced by our model, which takes into account both the primary+cascade synchrotron and SSC components. The fitting results are shown in Figure~\ref{LC}, with the corresponding model parameters summarized in Table~\ref{parameter sets}. As shown in the inset of the panel, the prompt gamma-ray emission precedes the onset of the afterglow phase. 
   
   \item GRB 090510. This is a bright, short, and hard burst, dominated by a matter component, with a redshift of $z = 0.903$ \citep{2010ApJ...709L.146D,2009GCN..9353....1R}. Its isotropic-equivalent gamma-ray energy in the prompt phase is $E_{\gamma,\rm iso} = 4.47^{+4.06}_{-3.77} \times 10^{52}\, \rm erg$ \citep{2011ApJ...730..141Z}. {\em Fermi}/LAT observations reveal highly energetic events \citep{2009Natur.462..331A, 2010MNRAS.409..226K}, and we incorporate the observed optical data (2 eV) into our analysis \citep{2010ApJ...709L.146D, 2010MNRAS.409..226K}. 
   Data are interpreted within the framework of central energy injection during the afterglow phase, with $E_{\rm in} = 2 \times 10^{52}~\mathrm{erg}$ and $t_{\tau} = 1613~\mathrm{s}$. The observed multi-wavelength lightcurves are well reproduced by the GRB afterglow model, taking into account both the primary+cascade synchrotron and SSC components. The results are presented in Figure~\ref{LC} and Table~\ref{parameter sets}. From the evolution of the observed data, the afterglow lightcurve subsequently enters the jet break stage at $\sim 2000$ seconds. We derived the jet opening angle of $0.77^\circ$, consistent with the $\sim 1^\circ$ estimate of \cite{2010MNRAS.409..226K} and the $\lesssim 1^\circ$ constraints from the numerical afterglow fitting of \cite{2011ApJ...733...22H}.
\end{itemize}

\section{Discussion}
\subsection{Magnetization Imprints of the Prompt Emission on the Afterglow}
The prompt and afterglow emissions of GRBs are generally believed to originate from distinct physical regions and processes. The prompt emission is typically produced by internal dissipation mechanisms within the ultra-relativistic jet.
For a matter-dominated fireball, the emission site is the internal shock radius $R_{\rm IS}\sim10^{13}-10^{14}\,\rm cm$ \citep{1994ApJ...430L..93R}. For a Poynting-flux-dominated outflow, a representative magnetic dissipation model is the Internal-collision-induced Magnetic Reconnection and Turbulence (ICMART) model, which predicts a prompt emission radius of $R_{\rm ICMART}\sim10^{15}-10^{16}\,\rm cm$ \citep{2011ApJ...726...90Z}. This phase is characterized by highly variable and intense gamma-ray and X-ray emission, reflecting the activity of the central engine and the jet dynamics. 
As shown in Figure~\ref{LC}, the GeV gamma-ray afterglows are detected at the late epoch of the prompt emission phase. The overlapping effect causes contamination from the late prompt emission to the afterglow data. 

We fit the time-integrated prompt gamma-ray spectra of GRBs 080916C and 090902B by using the joint {\em Fermi}/LAT and {\em Fermi}/GBM data. The spectrum of GRB 080916C is well fitted by an empirical Band function \citep{1993ApJ...413..281B}. The best-fit parameters are: low-energy photon index $\alpha= -0.98^{+0.04}_{-0.03}$, high-energy photon index $\beta = -2.20^{+0.03}_{-0.01}$, break energy $E_{\rm break} \sim 611.71^{+88.74}_{-102.01}$~keV, with a reduced chi-squared $\chi^2/\mathrm{dof} = 408/441$ (dof is the degrees of freedom).
The time-integrated prompt spectrum of GRB 090902B is well fitted by a Band function plus a power-law (PL) model, yielding $\chi^2/\mathrm{dof} = 682/414$. The Band component has $\alpha= -0.51\pm 0.03$, $\beta = -3.64^{+0.46}_{-0.24}$, and $E_{\rm break} = 500.89^{+22.88}_{-21.69}$~keV for the GBM data. The steep high-energy photon index agrees with a quasi-thermal origin analyzed by \cite{2011MNRAS.415.1663T} and \cite{2011ApJ...730..141Z}. Afterglow modeling gives similar electron distribution indices for GRBs 080916C and 090902B ($p = 2.4$ and $p = 2.3$, respectively), indicating that particle acceleration in the afterglow phase is not related to the properties of the prompt gamma-ray jets. We should also note that the PL component spans the keV-GeV band with $\Gamma = 1.94\pm0.01$, inconsistent with the prediction from the synchrotron radiation of an electron population with an index $p=2.3$.  

We estimate the electron magnetization of the afterglow jet with a ratio $\sigma_{B} = \epsilon_{B} / \epsilon_{e}$. Interestingly, $\sigma_B$ attains values of $4.9\times 10^{-4}$ and  $3.2\times 10^{-3}$ for GRBs 090902B and 090510, respectively, which are notably higher than those of GRBs 080916C ($ 5.9\times 10^{-5}$ ) and 130603B ( $3.2\times 10^{-5}$). This implies that in the prompt gamma-ray phase, a Poynting-flux-dominated GRB jet seems to have a low magnetization parameter in the afterglow. One possible explanation is that a Poynting-flux-dominated GRB jet has a higher magnetic dissipation efficiency than that of the matter-dominated GRB jet, leaving a weak seed magnetic field to the afterglow. We further investigate whether $\sigma_B$ is related to the prompt gamma-ray efficiency, defined as $\eta=E_{\gamma,\rm iso}/(E_{\gamma,\rm iso}+E_{\rm k, iso})$. Using the $E_{\rm k, iso}$ derived from our model fits, we obtain $\eta \simeq 33\% $ and $\eta \simeq 5\% $ for the Poynting-flux-dominated GRBs 080916C and 090510. It is $\eta \simeq 4\% $ and $\eta \simeq 11\% $ for the matter-dominated GRBs 090902B and 090510. No clear correlation between $\sigma_B$ and $\eta$ is observed among these four GRBs. 

The magnetic field strength in the comoving frame of the shocked fluid in the afterglow is given by $B' = ({32\pi \epsilon_B \Gamma^2 n m_p c^2})^{1/2}$, and therefore scales with both $\Gamma$ and $n$. 
We present the temporal evolution of the shock radius in the observer frame in the left panel of Figure~\ref{R(t)-B(t)}. The radial evolution profiles of the Poynting-flux-dominated long GRB 080916C and the matter-dominated long GRB 090902B exhibit remarkable similarities. For the short GRBs 090510 and 130603B, higher initial Lorentz factors lead to shorter deceleration timescales, indicating that the fireball reaches the deceleration stage earlier. 
The right panel of Figure~\ref{R(t)-B(t)} shows the temporal evolution of the magnetic field strength in the comoving frame. For long GRBs 080916C and 090902B, the magnetic field strength declines rapidly throughout the afterglow phase. The short GRBs show distinct magnetic field evolution characteristics: GRB 090510 shows a more gradual decline, while GRB 130603B maintains a relatively stable magnetic field strength of approximately $3\times10^{-2}$ Gauss up to 100 s, after which it drops sharply, similar to the behavior of GRB 090510. The reference parameter values are summarized in Table~\ref{parameter sets}.

\subsection{Electromagnetic cascade radiation in the Afterglow of GRB jets}
\label{sec:discussion}
We incorporate electromagnetic cascade radiation in our radiation physics model. To illustrate the contribution of this emission component in our fits to the SEDs of the four GRBs,  Figure~\ref{opacity and SEDs} shows the temporal evolution of the photon-photon opacity $\tau_{\gamma\gamma}$ for 0.1 TeV, 1 TeV, and 10 TeV photons in the comoving frame, and the predicted broadband SEDs.   
For the Poynting-flux-dominated long GRB 080916C, $\tau_{\gamma\gamma}$ for 10 TeV and 1 TeV photons exceeds unity during the very early afterglow, while 0.1 TeV photons escape after $\sim 5\,\rm s$. We present the SEDs over 60–80 s, when the prompt emission transitions to the afterglow. As shown in the middle panel of Figure~\ref{opacity and SEDs}, cascade synchrotron radiation dominates the radio to optical bands, consistent with significant $\gamma\gamma$ absorption. At the late afterglow stage of $(6\text{--}7)\times10^{5}$ s, the cascade process is no longer significant (right panel). The observational data for each time slice are incorporated, with Galactic extinction properly accounted for. 
For the matter-dominated long GRB 090902B, 0.1 TeV and 1 TeV photons escape during the very early afterglow phase, while 10 TeV photons are likely extinguished via $\gamma\gamma$ absorption. Subsequently, weak cascade emission appears in the radio-to-optical bands over 40–60 s. During the later afterglow phase of $(1\text{--}2)\times10^{5}$ s, secondary radiation becomes completely negligible. For the short GRBs 130603B/090510, $\gamma\gamma$ absorption is insignificant. Thus, the entire spectral range is dominated by primary synchrotron and SSC processes. The results in Figure~\ref{opacity and SEDs} and the reference parameter values are summarized in Table~\ref{parameter sets}.

The left panel of Figure~\ref{Gamma} shows the temporal evolution of $\gamma_{c}$, $\gamma_{m}$, and $\gamma_{\rm max}$, which characterize the energy distribution of electrons accelerated in the GRB jet. The right panel presents the evolution of $Y(\gamma_{c})$ and $Y(\gamma_{m})$ as functions of time, where $Y$ quantifies the relative importance of inverse Compton scattering versus synchrotron cooling for electrons with $\gamma_{c}$ and $\gamma_{m}$, respectively. 
For the Poynting-flux-dominated long GRB 080916C, the radiating electrons initially reside in the fast cooling regime, transitioning to the slow cooling regime after $\sim 30\rm\,s$. Correspondingly, $Y(\gamma_{c})$ rapidly decreases to unity around 30 s, after which $Y(\gamma_{m})$ becomes the dominant parameter. For the matter-dominated long GRB 090902B, the electrons start in the fast cooling regime and transition to the slow cooling after $\sim 20\rm\,s$. Here, $Y(\gamma_{c})$ rises to unity at $\sim 20\rm\,s$ and then declines rapidly. For the short GRBs 130603B/ 090510, the electrons remain in the slow cooling regime throughout the afterglow phase, with $Y(\gamma_{m})$ consistently governing the cooling process.

\section{Summary}
In this work, we systematically investigate the impact of jet composition (Poynting-flux-dominated vs. matter-dominated) on the GRB afterglow emission using representative long GRBs (080916C, 090902B) and short GRBs (130603B, 090510). Within the framework of primary+cascade synchrotron and SSC  processes incorporating $\gamma\gamma$ absorption, our model successfully reproduced the observed multi-band afterglow lightcurves and SEDs with reasonable physical parameters. Our results show that the magnetic and baryonic energy components are efficiently converted into forward shock kinetic energy, rendering the initial jet composition a negligible factor in shaping the afterglow emission. These findings provide important insights into energy conversion processes in GRB jets and demonstrate the universality of afterglow radiation mechanisms across different jet compositions.

\acknowledgments
We appreciate the insightful remarks of the referee that helped to improve the manuscript. This work is supported by the National Natural Science Foundation of China (grant No. 12203015) and the Guizhou Normal University startup financial support program (grant No. GZNUD[2023]). Yu Gan acknowledges support from the College Student Innovation and Entrepreneurship Training Program (grant No. s202310663006).
  
\clearpage
\begin{table}[htbp]
    \centering
    \caption{Results of the Theoretical Fits to Multi-Wavelength Afterglow Lightcurves with External Shock Models by Considering Electromagnetic cascade emission}
    \setlength{\tabcolsep}{4pt}
    \begin{tabular}{lccccccccccc}
        \toprule
   GRBs & $z$ & $T_{\rm 90}$ & Composition & $\log(E_{\rm k,iso})$ & $\log(n_0)$ & $\log(A_{*})$ & $\log(\epsilon_e)$ & $\log(\epsilon_B)$ & $p$ & $\log(\Gamma_0)$ & $\log(\theta_j)$ \\
     &  & (s) & &($\rm erg$) &  ($\mathrm{cm}^{-3}$) &  &  &  &  &  & ($\rm rad$)\\
    \midrule
    080916C &   4.35 &   66 & Poynting-flux  & $55.06^{+0.47}_{-0.65}$ & $-$ & $-1.83^{+0.63}_{-0.68}$ & $-0.92^{+0.30}_{-0.45}$ &$-3.91^{+0.67}_{-0.72}$ & 2.4 & $2.97^{+0.25}_{-0.19}$ & $-$\\
    090902B & 1.822 & 21.9 &    matter  & $54.81^{+0.30}_{-0.19}$ & $-$ & $-1.75^{+0.26}_{-0.34}$ & $-1.19^{+0.20}_{-0.20}$ & $-3.74^{+0.43}_{-0.45}$ & 2.3 & $3.06^{+0.16}_{-0.21}$ & $-$\\
    \midrule
    130603B & 0.3568 & 0.18 & Poynting-flux &  $52.76^{+0.47}_{-0.31}$ & $-2.52^{+0.70}_{-0.52}$ & $-$ & $-0.76^{+0.17}_{-0.19}$ & $-4.58^{+0.95}_{-0.98}$ & 2.4 & $2.28^{+0.08}_{-0.12}$ & $-$\\   
    090510 &  0.903 & 0.3 &  matter & $52.60^{+0.15}_{-0.08}$ &  $-3.58^{+0.21}_{-0.29}$ & $-$ & $-0.55^{+0.02}_{-0.05}$ & $-3.10^{+0.08}_{-0.15}$ & 2.4 & $3.42^{+0.11}_{-0.16}$ & $-1.87^{+0.04}_{-0.05}$\\
    \bottomrule
    \end{tabular}
    \label{parameter sets}
\end{table}

\begin{figure}[htbp]
    \centering
    \includegraphics[scale=0.4]{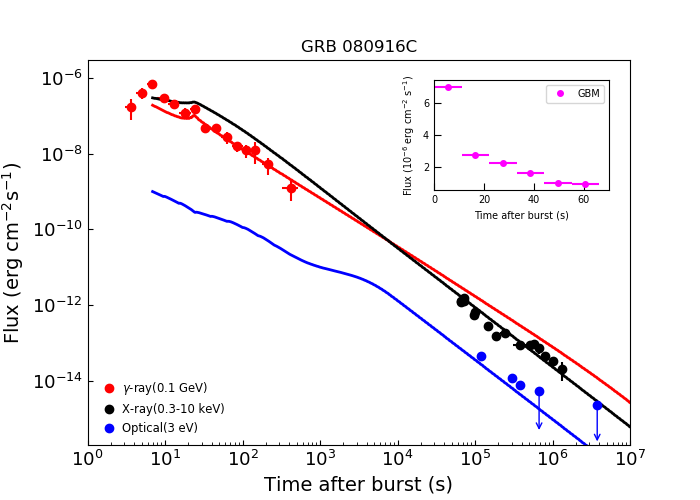}
    \includegraphics[width=0.3\textwidth]{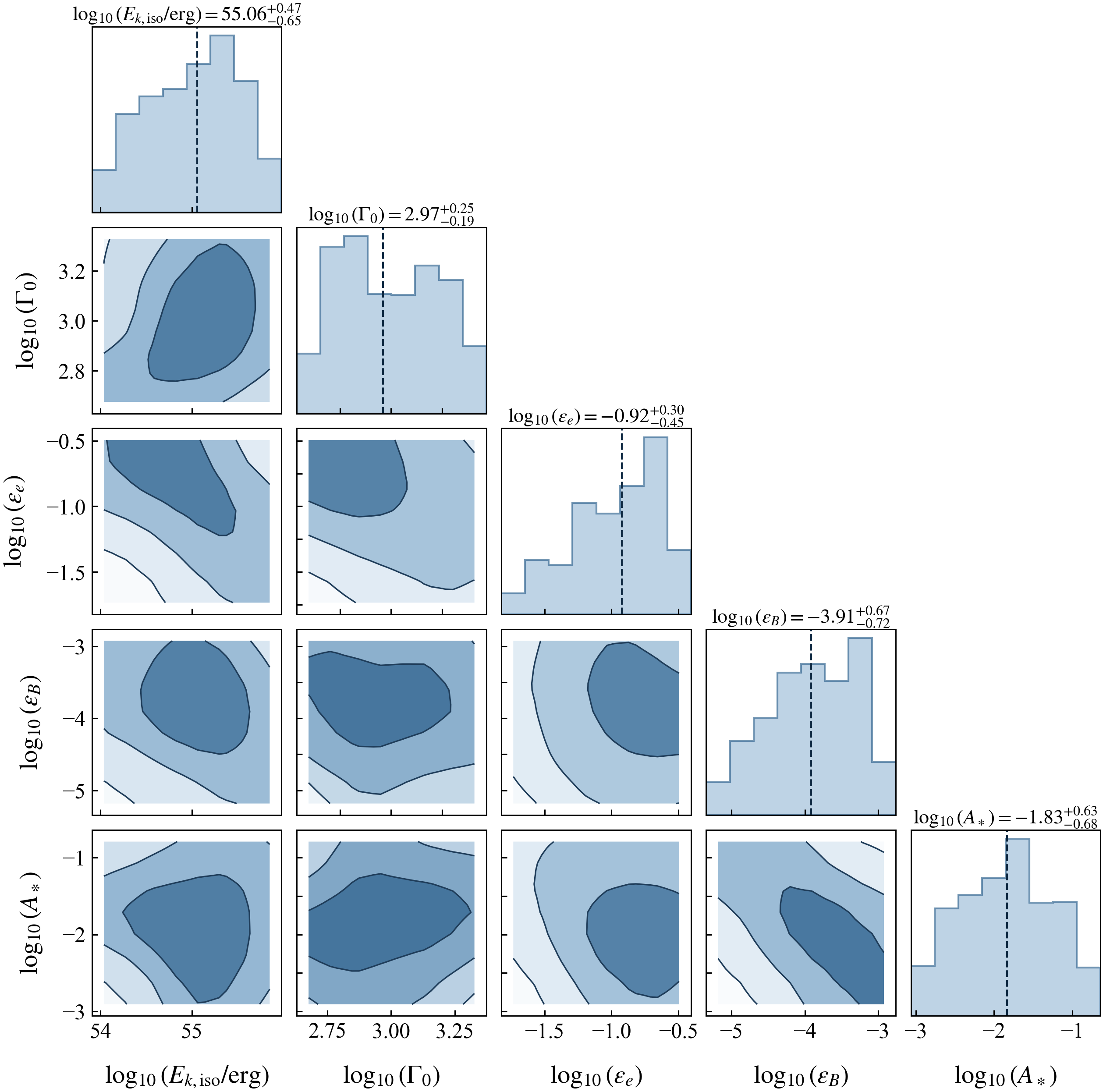}
    \includegraphics[scale=0.4]{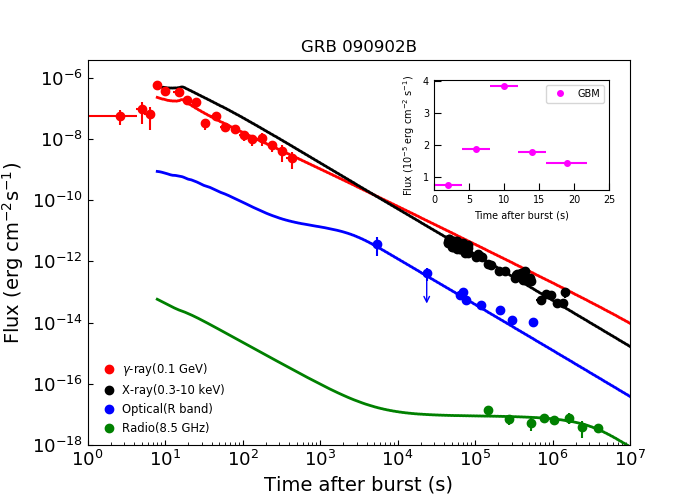}
    \includegraphics[width=0.3\textwidth]{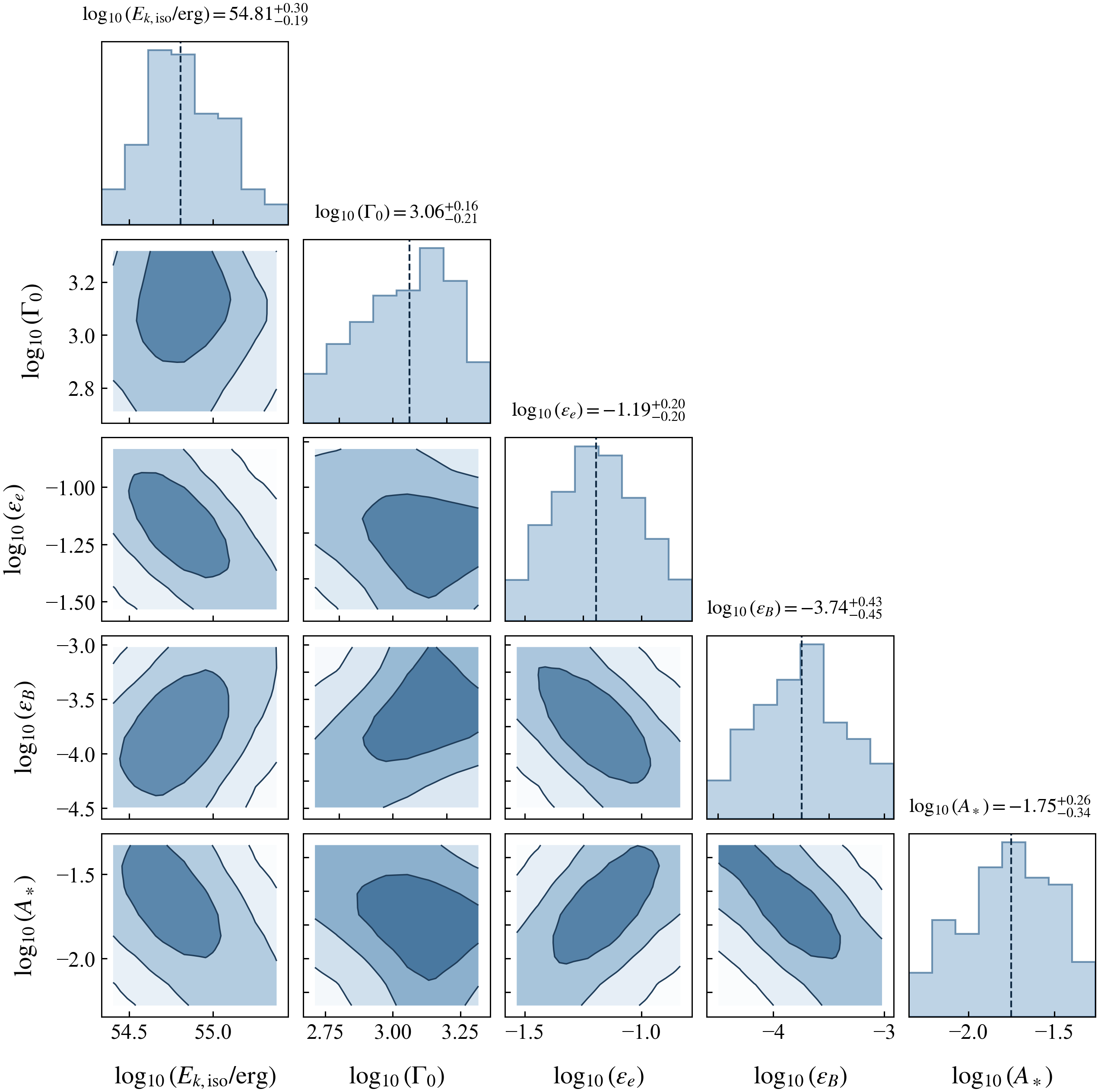}
    \includegraphics[scale=0.4]{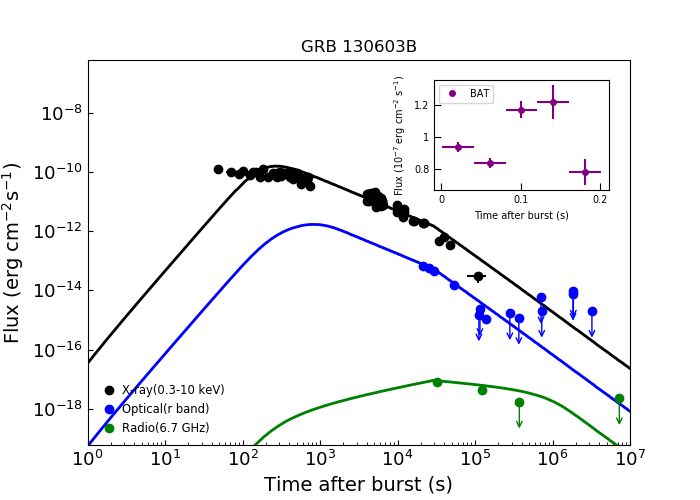}
    \includegraphics[width=0.3\textwidth]{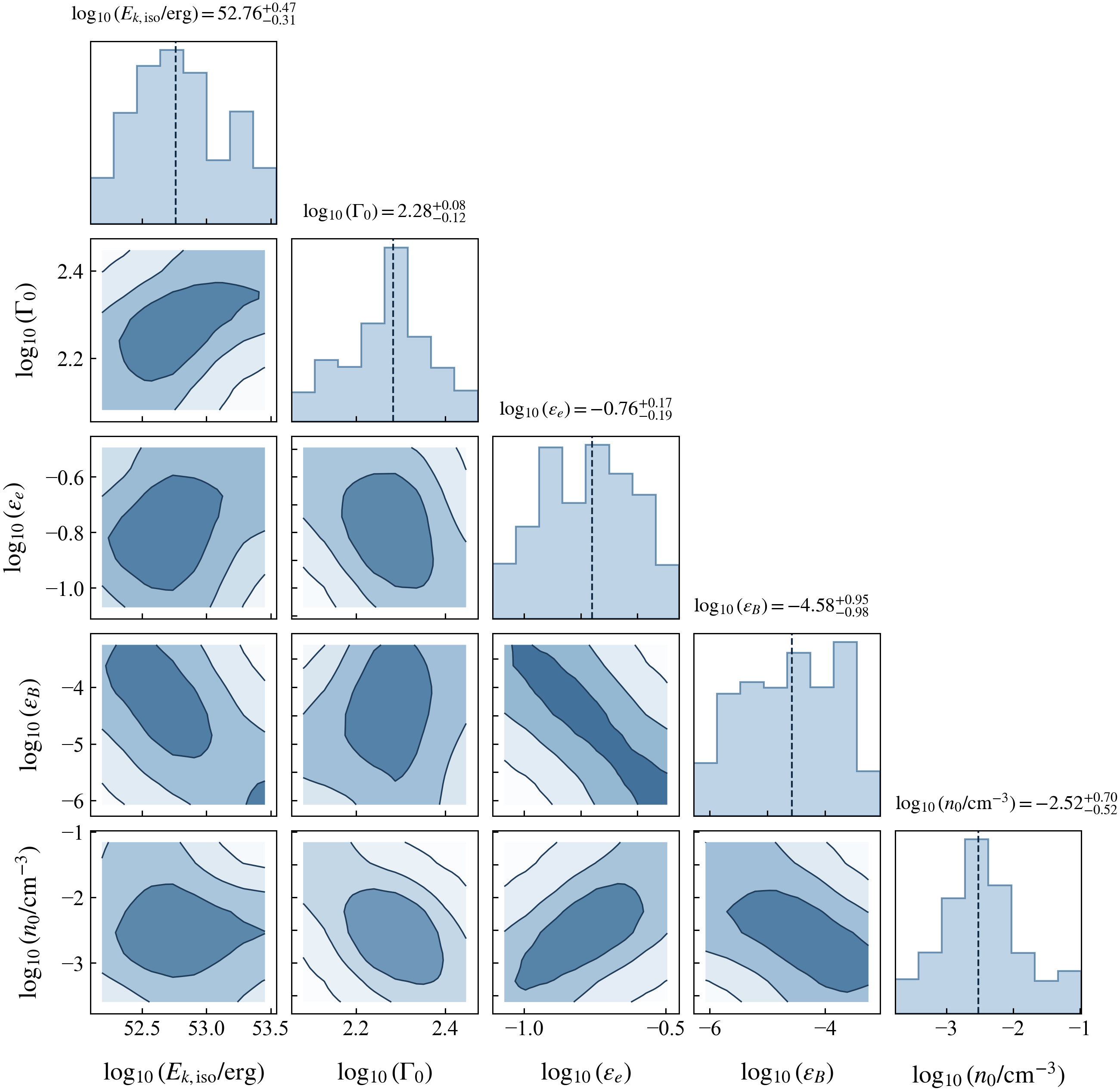}
    \includegraphics[scale=0.4]{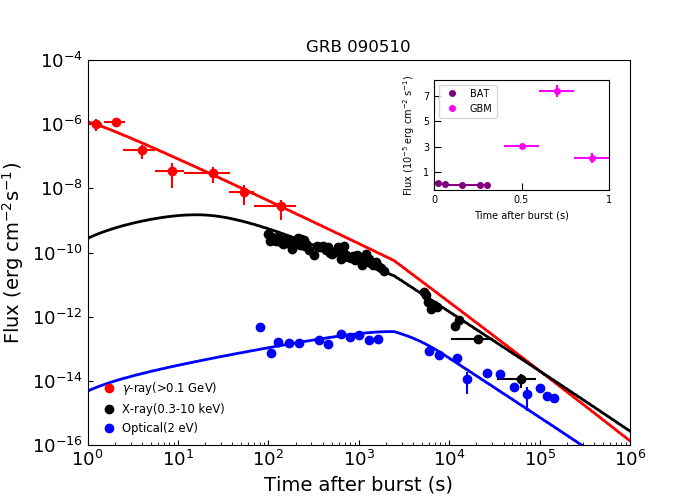}
    \includegraphics[width=0.3\textwidth]{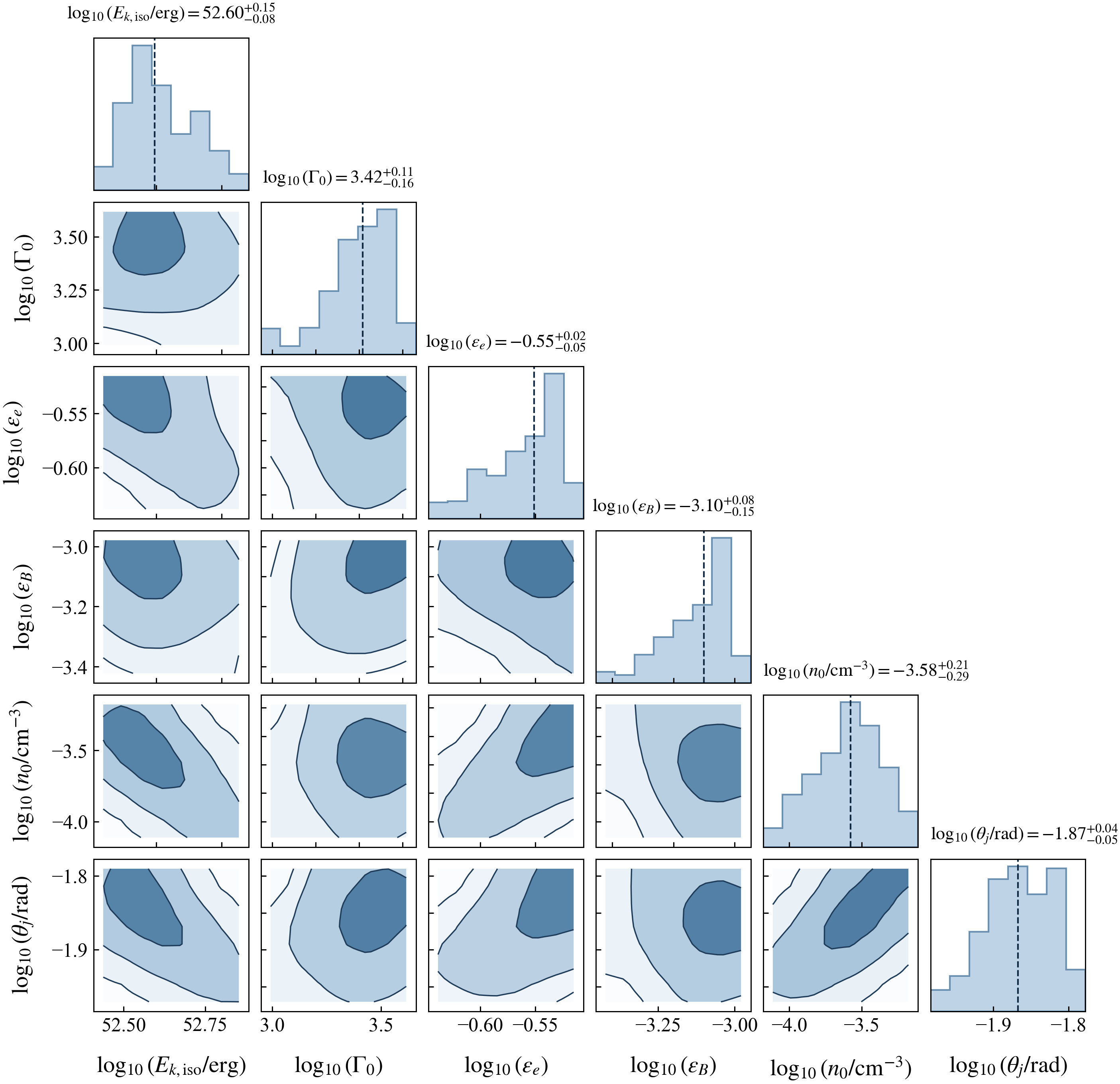}
    \caption{Left panel: the observed multi-wavelength lightcurves (dots) and the best theoretical fits (solid lines) of long GRBs 080916C/090902B and short GRBs 130603B/090510 with our GRB afterglow model. The solid lines represent the sum of the emission from the primary synchrotron+SSC components, and cascade synchrotron+SSC processes by considering KN effect. The parameter values used are shown in Table~\ref{parameter sets}. The inset of the panel shows prompt gamma-ray lightcurves in the linear timescale. The vertical gray dashed lines separates the prompt emission from the afterglow for long GRBs 080916C/090902B. Right panel: posterior distributions of our afterglow model parameters derived from MCMC simulations for GRBs 080916C, 090902B, 130603B, and 090510, respectively.}  \label{LC}
\end{figure}

\begin{figure}[ht]	
    \centering	
    \includegraphics[scale=0.5]{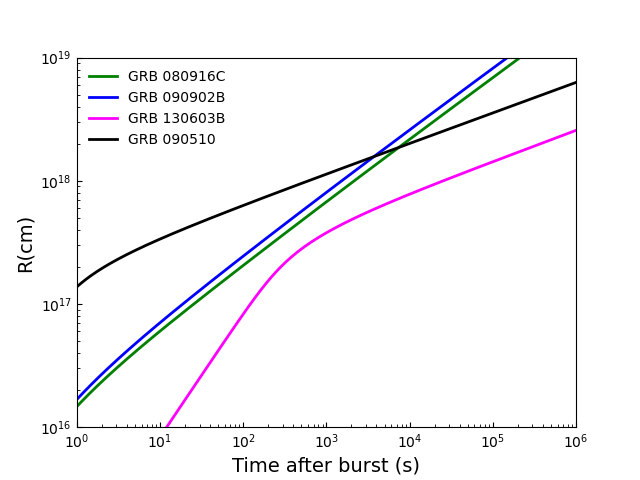}
    \includegraphics[scale=0.5]{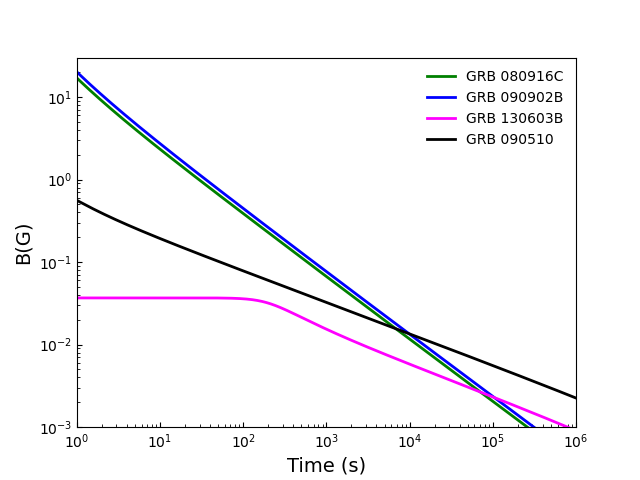}
    \caption{Radius and Magnetic field strength as a function of time. The parameter values used are the same as those in Figure~\ref{LC}.}
    \label{R(t)-B(t)}
\end{figure}

\begin{figure}[ht]	
    \centering
    \includegraphics[scale=0.333]{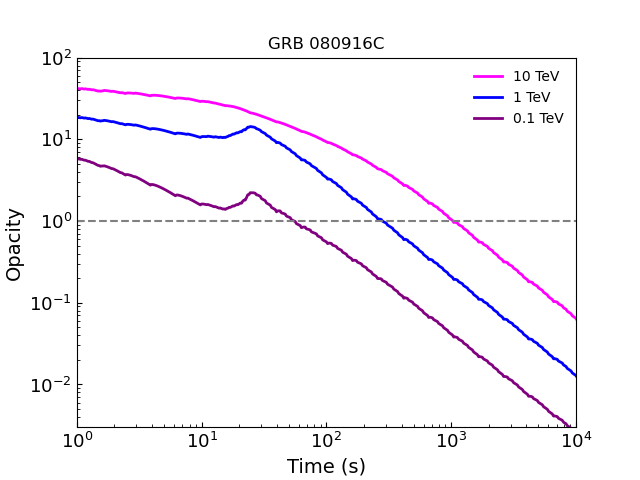}
    \includegraphics[scale=0.333]{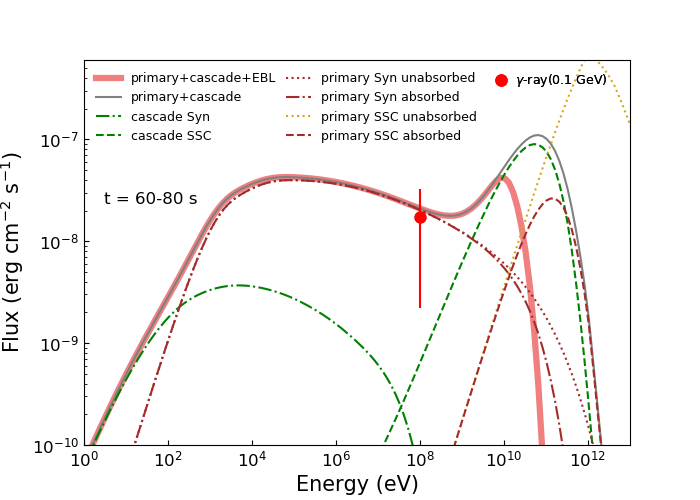}
    \includegraphics[scale=0.333]{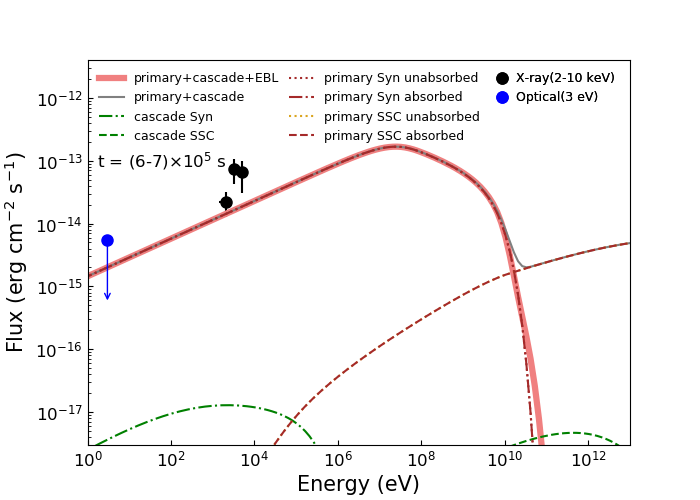}
    \includegraphics[scale=0.333]{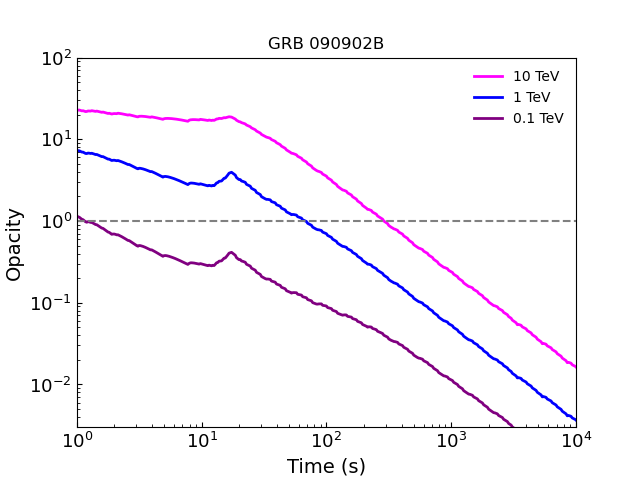}
    \includegraphics[scale=0.333]
    {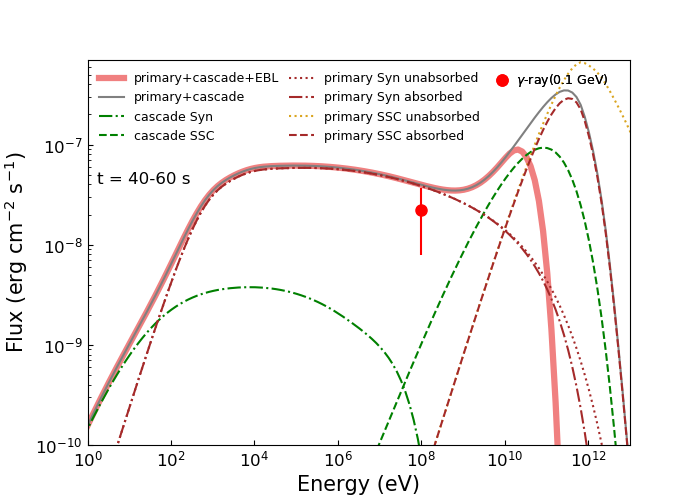}
    \includegraphics[scale=0.333]{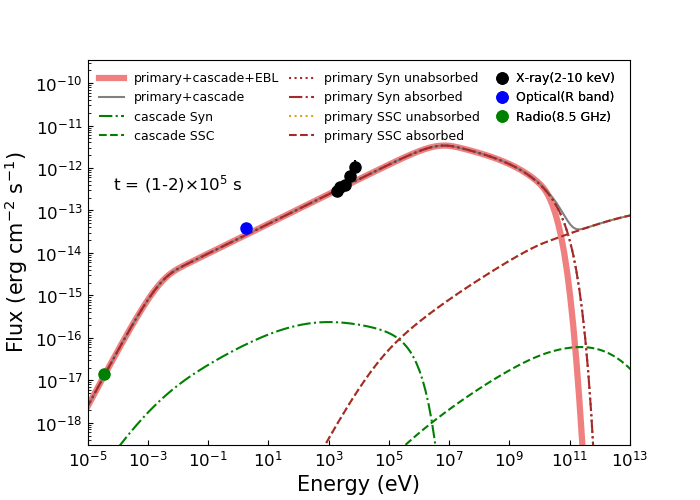}
    \includegraphics[scale=0.333]{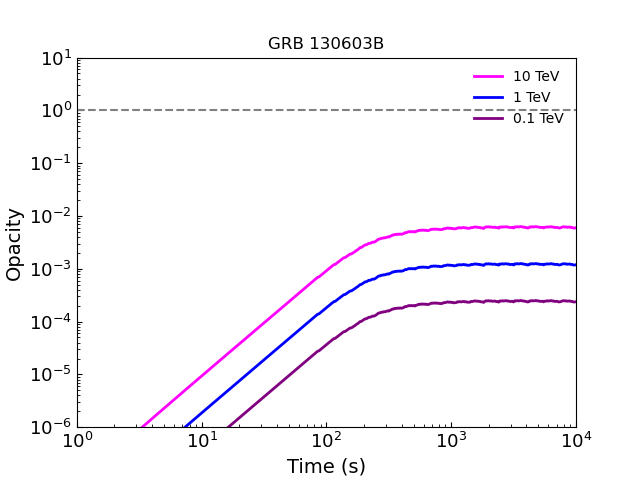}
    \includegraphics[scale=0.333]{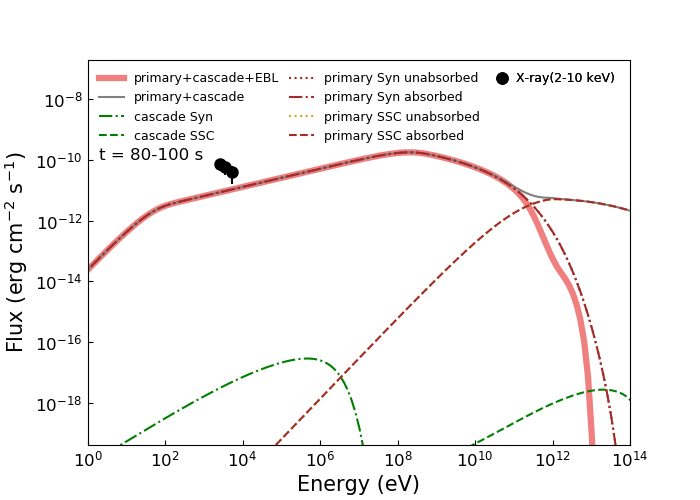}
    \includegraphics[scale=0.333]{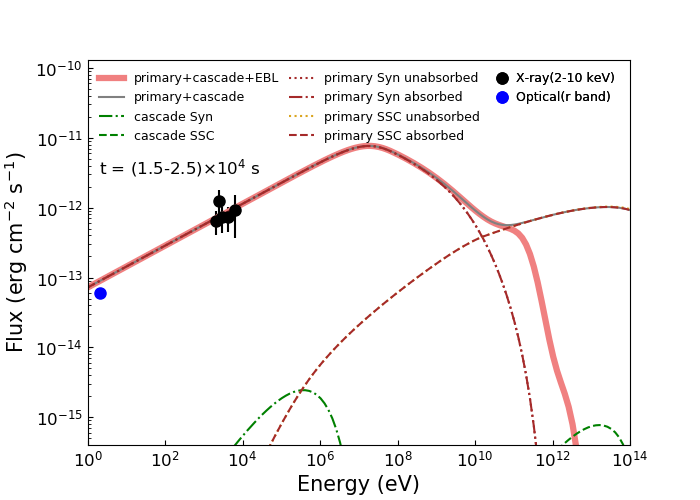}
    \includegraphics[scale=0.333]{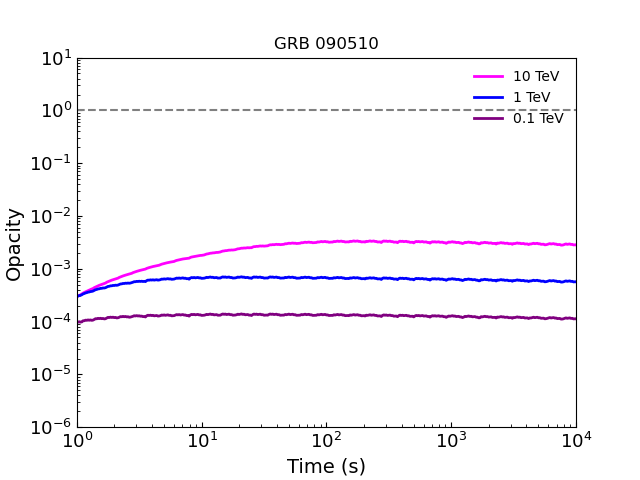}
    \includegraphics[scale=0.333]{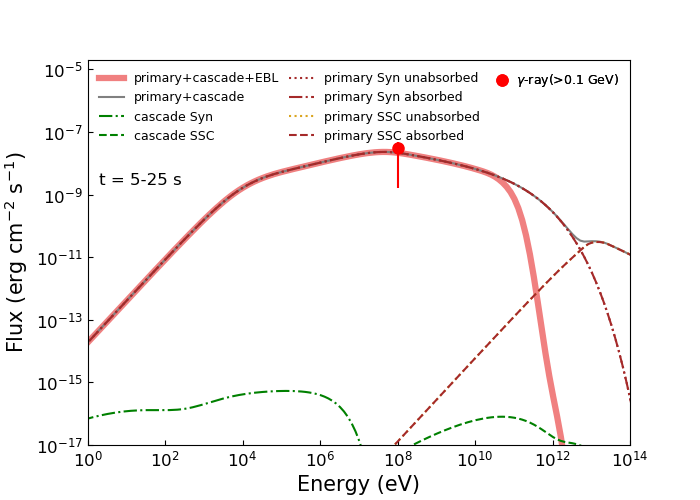}
    \includegraphics[scale=0.333]{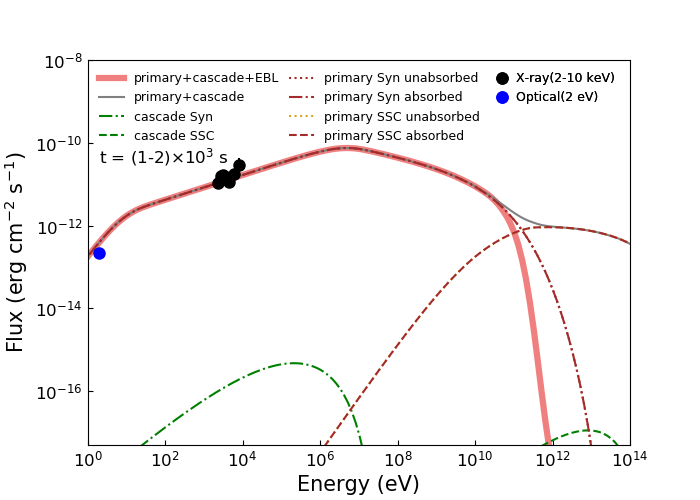}
    \caption{Left panel: temporal evolution of the opacity of 0.1 TeV, 1 TeV and 10 TeV gamma-ray photons within the afterglow jet. The gray dashed lines represent $\tau_{\gamma\gamma}=1$. Middle/Right panel: the broadband SEDs of afterglows after the GRB trigger of Poynting-flux-dominated GRBs 080916C/130603B and Matter-dominated GRBs 090902B/090510, respectively. The solid orange lines represent the sum of the emission from the primary synchrotron radiation (the dash-dot brown lines), primary SSC process (the dashed brown lines), the cascade synchrotron radiation (the dash-dot green lines), and the cascade SSC process (the dashed green lines) with considering EBL absorption. The dotted brown lines and dotted yellow lines represent the emission from the primary synchrotron emission and primary SSC process without considering the $\gamma\gamma$ absorbed effect, respectively. The 0.3–2 keV band is excluded from the spectral fitting due to heavy neutral hydrogen absorption. The parameter values used are the same as those in Figure~\ref{LC}.}
    \label{opacity and SEDs}
\end{figure}

\begin{figure}[ht]	
    \centering	
    \includegraphics[scale=0.49]{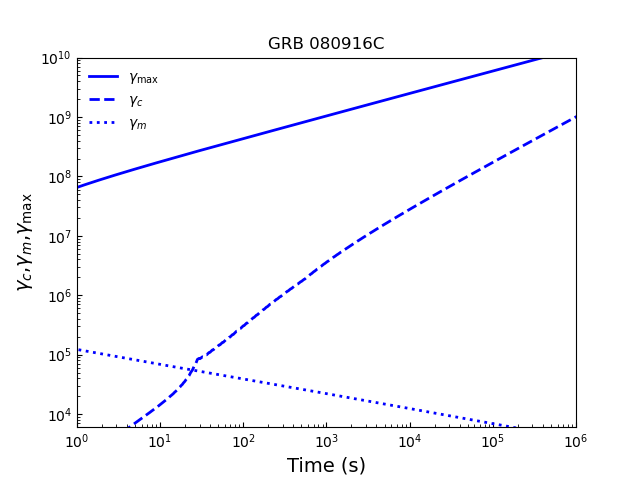}
    \includegraphics[scale=0.49]{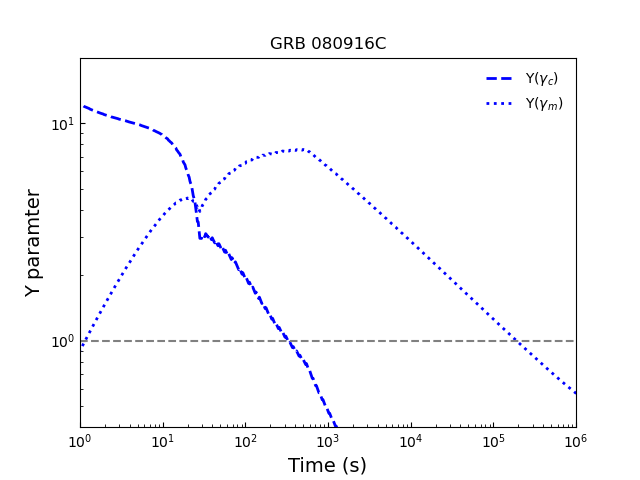}
    \includegraphics[scale=0.49]{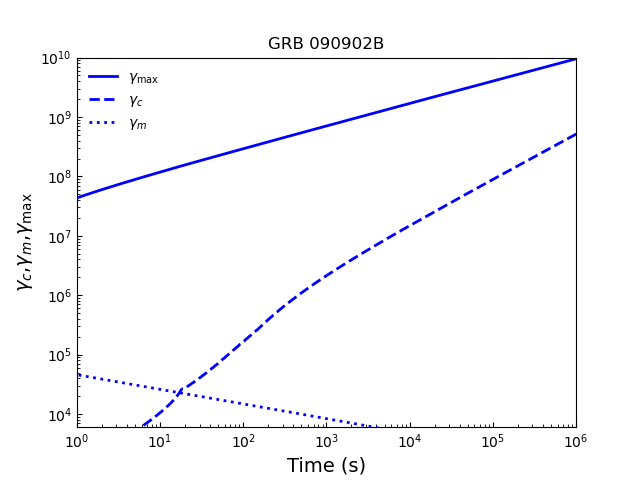}
    \includegraphics[scale=0.49]{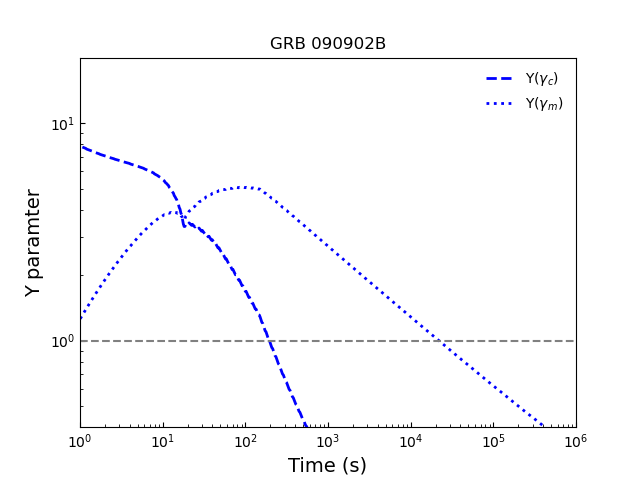}
    \includegraphics[scale=0.49]{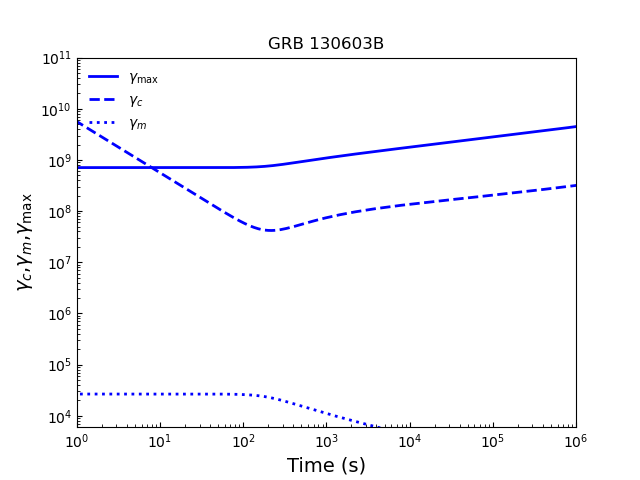}
    \includegraphics[scale=0.49]{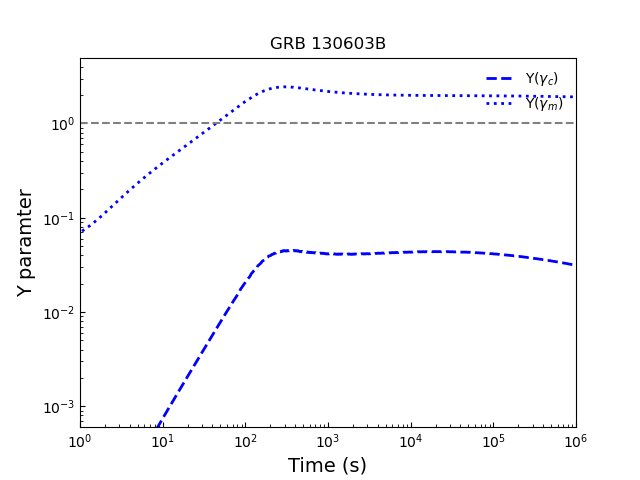}
    \includegraphics[scale=0.49]{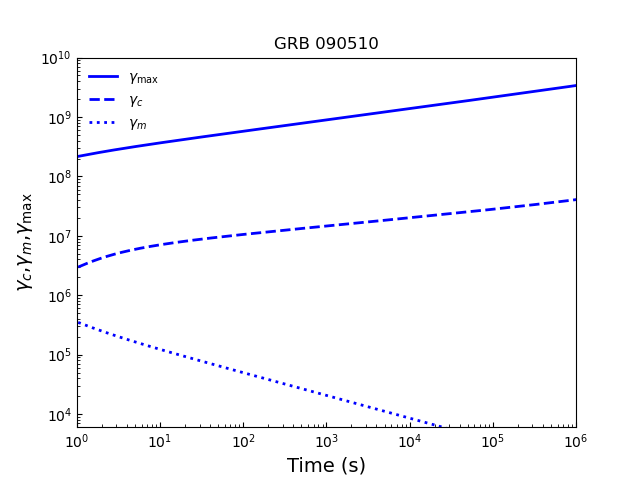}
    \includegraphics[scale=0.49]{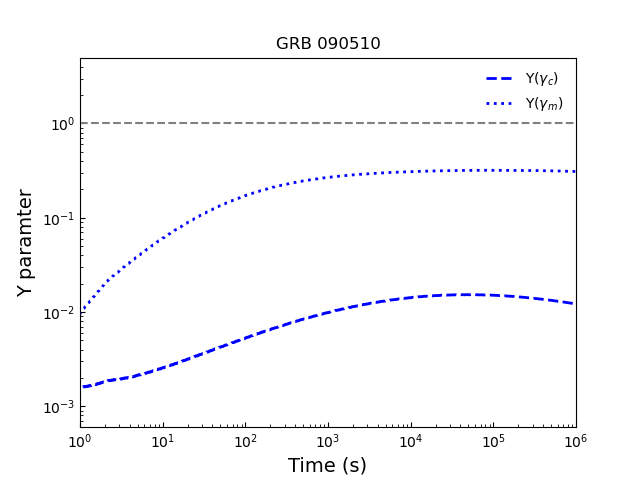}
    \caption{Left panel: the values of $\gamma_{c}$, $\gamma_{m}$, and $\gamma_{\mathrm{max}}$ as a function of time. Right panel: Compton parameters $Y(\gamma_{c})$ and $Y(\gamma_{m})$ as a function of time. The parameter values used are the same as that in Figure~\ref{LC}.}
    \label{Gamma}
\end{figure}

\end{document}